# Are topic-specific search term, journal name and author name recommendations relevant for researchers?

Philipp Mayr
GESIS Leibniz Institute for the Social
Sciences
Unter Sachsenhausen 6-8
50667 Cologne, Germany
+49 (0)221 – 47694 - 0
philipp.mayr@gesis.org

## ABSTRACT
In this paper we describe a case study where researchers in the social sciences (n=19) assess topical relevance for controlled search terms, journal names and author names which have been compiled automatically by bibliometric-enhanced information retrieval (IR) services. We call these bibliometric-enhanced IR services Search Term Recommender (STR), Journal Name Recommender (JNR) and Author Name Recommender (ANR) in this paper. The researchers in our study (practitioners, PhD students and postdocs) were asked to assess the top n pre-processed recommendations from each recommender for specific research topics which have been named by them in an interview before the experiment. Our results show clearly that the presented search term, journal name and author name recommendations are highly relevant to the researchers' topic and can easily be integrated for search in Digital Libraries. The average precision for top ranked recommendations is 0.75 for author names, 0.74 for search terms and 0.73 for journal names. The relevance distribution differs largely across topics and researcher types. Practitioners seem to favor author name recommendations while postdocs have rated author name recommendations the lowest. In the experiment the small postdoc group (n=3) favor journal name recommendations.

## Categories and Subject Descriptors
H.3.3 [**Information Search and Retrieval**]: Relevance feedback, Retrieval models, Selection process.

## General Terms
Measurement, Experimentation, Human Factors.

## Keywords
Recommendation services, bibliometric-enhanced IR, Bradfordizing, co-word analysis, author centrality, relevance assessment.

## 1. INTRODUCTION
In metadata-driven digital libraries (DL) typically three major difficulties arise: (1) the vagueness between search and indexing terms, (2) the information overload by the amount of result records obtained by information retrieval (IR) systems, and (3) the problem that pure term frequency based rankings, such as term frequency – inverse document frequency (tf-idf), provide results that often do not meet user needs.

Search term suggestion or other domain-specific recommendation modules can help users – especially in the social sciences and humanities [3, 4] – to formulate their queries by mapping their personal vocabularies onto the highly specialized vocabulary of a digital library. A recent overview of recommender systems in DL can be found in [4]. We will present an approach to utilize bibliometric-enhanced information retrieval services [7] as enhanced search stratagems [1] and recommendation services within a scholarly IR environment [12, 13]. The bibliometric-enhanced information retrieval services can easily be implemented within scholarly information portals (see e.g. our interactive web-based prototype[1]). We assume that a user's search should improve by using these recommendation services when interacting with a scientific information system [15]. Even though, researchers may find expansion terms presented by a recommendation service useful, but the relevance of these recommendations in a real search task is still an open question (see also the experiences with the Okapi system [17]).

The idea of the paper is to evaluate if topic-specific recommendation services which provide search related thesaurus term, journal name and author name suggestions are accepted by researchers. This paper will shortly introduce three bibliometric-enhanced information retrieval services: (1) co-word analysis and the derived concept of Search Term Recommendation (STR), (2) coreness of journals and the derived Journal Name Recommender (JNR) and (3) centrality of authors and the derived Author Name Recommender (ANR). The basic concepts and an evaluation of the top-ranked recommendations are presented.

## 2. MODELS FOR INFORMATION RETRIEVAL ENHANCEMENT
The standard model of IR is the tf-idf model which proposes a text-based relevance ranking [6]. As tf-idf is text-based it assigns a weight to term t in document d which is influenced by different occurrences of t and d. Variations of the basis term weighing process have been proposed, like normalization of document length or by scaling the tf values but the basic assumption stays the same. The following recommendation services are outlined very shortly. More details can be found in [8].

---
[1] http://www.gesis.org/beta/prototypen/irm/

## 2.1 Search Term Recommendation

Search Term Recommenders (STR) are an approach to compensate the long known language problem in IR [2, 10, 4]: Searching an information system a user has to come up with the "correct" query terms so that they best match the document language to get an appropriate result. In this paper STR are based on statistical co-word analysis and generate associations between free terms (i.e. from title or abstract) and controlled terms (i.e. from a thesaurus) which are used during a professional indexing of the documents. The co-word analysis implies a semantic association between the free and the controlled terms. The more often terms co-occur in the text the more likely it is that they share a semantic relation. In our setup we use STR for search term recommendation where the original topical query of the researcher is expanded with semantically "near" terms from a controlled vocabulary.

## 2.2 Recommending Journal Names

Journals play an important role in the scientific communication process. They appear periodically, they are topically focused, they have established standards of quality control and often they are involved in the academic reward system. Metrics like the famous impact factor are aggregated on the journal level. In some disciplines journals are the main place for a scientific community to communicate and discuss new research results. In addition, journals or better journal names play an important role in the search process (see the famous search stratagem "journal run" [13, 9]).

The underlying mechanism for recommending journal names (JNR) in this paper is called Bradfordizing [15]. Bradfordizing is an alternative mechanism to re-rank result lists according to core journals to bypass the problem of very large and unstructured result sets. The approach of Bradfordizing is to use characteristic concentration effects (Bradford's law of scattering) that appear typically in journal literature. Bradfordizing defines different zones of documents which are based on the frequency counts in a given document set. Documents in core journals – journals which publish frequently on a topic – are ranked higher than documents which were published in journals from the subsequent zones. In IR a positive effect on the search result can be assumed in favor of documents from core journals [8]. Bradfordizing is implemented as "journal productivity" in the digital library sowiport[2].

**Figure 1. Recommending author names in our retrieval prototype. Example search term "luhmann" and highly associated author names (central authors).**

---

[2] http://sowiport.gesis.org/

## 2.3 Recommending Author Names

Collaboration in science is mainly represented by co-authorships between two or more authors who write a publication together. Transferred to a whole community, co-authorships form a co-authorship network reflecting the overall collaboration structure of a community.

The underlying mechanism for recommending author names (ANR) in this paper is the author centrality measure betweenness. We propose to use author centrality for re-ranking result sets. Here the concept of centrality in a network of authors is an additional approach for the problem of large and unstructured result sets. The intention behind this ranking model is to make use of knowledge about the interaction and cooperation behavior in special fields of research. The (social) status and strategic position of a person in a scientific community is used, too. The model is based on a network analytical view on a field of research and differs greatly from conventional text-oriented ranking methods like tf-idf. A concrete criterion of relevance in this model is the centrality of authors from retrieved publications in a co-authorship network. The model calculates a co-authorship network based on the result set to a specific query. Centrality of each single author in this network is calculated by applying the betweenness measure and the documents in the result set are ranked according to the betweenness of their authors so that publications with very central authors are ranked higher in the result list [8]. From a recent study we know that many users are searching DL with author names. In addition, author name recommendations basing on author centrality can successfully be used as query expansion mechanism [11].

## 2.4 Implementation

All proposed models are implemented in a live information system using (1) the Solr search engine, (2) Grails Web framework to demonstrate the general feasibility of the approaches. Both Bradfordizing and author centrality as re-rank mechanisms are implemented as plugins to the open source web framework Grails. Grails is the glue to combine the different modules and to offer an interactive web-based prototype. In general these retrieval services can be applied in different query phases. In a typical search scenario a user first formulates his query, which can then be enriched by a STR that adds controlled descriptors from the corresponding document language to the query. With this new query a search in a database can be triggered. The search returns a result set which can be re-ranked using either Bradfordizing or author centrality (see Figure 1). Since search is an iterative procedure, this workflow can be repeated many times untill the expected result set is retrieved.

In the following we will describe a case study with researchers using the recommendation services STR, JNR and ANR to find search terms, journal names and author names relevant to their research topics.

## 3. ASSESSMENT STUDY

The assessment study involved 19 researchers in the social sciences who agreed to name one or two of their research topics and take part in a short online assessment exercise. We have recruited the researchers (seniors, research staff and PhD candidates) via email and telephone and they were asked to qualify their primary research topic in the form of 1-3 typical search terms which they would typically enter in a search task. These search terms have been operationalized by us into a valid query for our prototype together with an individualized login for the single researcher. Individualized assessment accounts were sent to the researchers via email for each topic and contained a link to the online assessment tool and a short description how to evaluate the recommendations. All researchers were asked to judge topical relevance of each recommendation in relationship to their research topic (binary assessments). Graded assessments would be an alternative which will be considered in future work.

The following 23 topics were evaluated by the researchers: [east Europe; urban sociology; equal treatment; data quality; interviewer error; higher education research; evaluation research; information science; political sociology; party democracy; data quality (2), party system; factor structure; nonresponse; ecology; industrial sociology; sociology of culture; theory of action; atypical employment; lifestyle; Europeanization; survey design; societal change in the newly-formed German states].

## 4. EVALUATION

In the following we describe the evaluation of the recorded assessments. We calculated average precision P(av) for each recommender service. The precision $P$ of each service was calculated by

$$P = \frac{|r|}{|r+nr|}$$

for each topic, where $|r|$ is the number of all relevant assessed recommendations and $|r+nr|$ is the number of all assessed recommendations (relevant and not relevant).

We intended to keep the assessment exercise for the researchers very short and hence we limited the list of recommendations of each service to a maximum of 5 controlled terms, journal names and author names. According to this restriction we decided to calculate P@1, P@2, P@4 for each service. In very rare cases one recommendation service generated just one or two recommendations.

## 5. RESULTS

In sum 19 researchers assessed 23 topics in the online assessment study. This resulted in a total of 95 STR, 111 JNR and 107 ANR assessments (see Table 1). In average the researchers assessed 4.1 search term, 4.8 journal name and 4.6 author name recommendations per topic.

**Table 1. Statistics of the assessment study**

| Researchers | Topics | STR A. | JNR A. | ANR A. |
|---|---|---|---|---|
| 19 | 23 | 95 | 111 | 107 |

Table 2 shows the evaluation results of all STR, JNR and ANR assessments. For this case study we did no statistically testing like t-test or Wilcoxon because of our very small sample. The following results should be read as plausibility tests without any statistical significance. We just intend to demonstrate here the indicative relevance of this kind of recommender systems for scholarly search systems.

**Table 2. Evaluation of the assessments. Average precision, P@1, P@2 and P@4 for recommendation from STR, JNR and ANR**

|  | STR | JNR | ANR |
|---|---|---|---|
| **P(av)** | **0.743** | 0.728 | **0.749** |
| **P@1** | **0.957** | 0.826 | **0.957** |
| **P@2** | 0.826 | **0.848** | **0.864** |
| **P@4** | 0.750 | 0,726 | 0.750 |

We can see that the average precision P(av) of ANR (0.749) and STR (0.743) is slightly better than JNR (0.728). Consulting the P@1 measures ANR and STR are clearly better the JNR. That means that the first recommended author name or search term is rated more often relevant than the first journal name in a list of 4 or 5 recommendations. Surprisingly JNR (0.848) is slightly better than STR (0.826) in P@2. If we look at the last row (P@4) in Table 2 we can see that all three recommendation services move closer together when more recommendations are assessed.

Table 3 shows the average precision P(av) of STR, JNR and ANR for our three different researcher types (practitioners, PhD students and postdocs). From the 19 researchers in our user study we could group 8 researchers into the practitioners group (mostly information professionals without PhD), 8 PhD students which had 1-4 years research experience and a small group of 3 postdocs with 4 and more years of research experience. We can see clearly that the author name recommendations are rated highest by the practitioners (see P(av) of ANR = 0.836). Surprisingly the postdocs have evaluated ANR much lower than the other two groups (see P(av) of ANR = 0.467). In the experiment postdocs favor journal name recommendations. PhD students rate all three recommenders more or less the same.

**Table 3. Evaluation of different researcher types. Average precision for recommendation from STR, JNR and ANR**

|  | STR | JNR | ANR |
|---|---|---|---|
| **P(av) Practitioners** (N=8) | 0.727 | 0.709 | **0.836** |
| **P(av) PhD students** (N=8) | 0.742 | 0.719 | 0.737 |
| **P(av) Postdocs** (N=3) | 0.750 | **0.800** | 0.467 |

## 6. CONCLUSION

In this small case study typical researchers in the social sciences are confronted with specific recommendations which were calculated on the fly on the basis of researchers' research topics. Looking at the precision values two important insights need to be mentioned: (1) precision values of recommendations from STR, JNR and ANR are close together on a very high level (P(av) is close to 0.75) and (2) each service retrieved a disjoint set of relevant recommendations. The different services each favor quite other – but still relevant – recommendations and relevance distribution differs largely across topics and researchers.

A distinction between researcher types shows that practitioners are favoring author name recommendations (ANR) while postdocs are favoring journal name recommendations precompiled by our recommender services. This can be an artifact due to the small size of the postdoc group but this is also plausible. In terms of research topics author names typically are more distinctive than journal names. An experienced researcher (e.g. postdoc) who is familiar with an authors' work can quickly rate an authors' name relevance for a specific topic. In this context journal names are not that problematic because they are published widely on different topics. This seems to be the case in our small sample (see third row in Table 3). PhD students who typically are unexperienced find all recommendations (terms, author names, journal names) helpful (see second row in Table 3).

The proposed models and derived recommendation services open up new viewpoints on the scientific knowledge space and also provide an alternative framework to structure and search domain-specific retrieval systems [8, 9]. In sum, this case study presents limited results but shows clearly that bibliometric-enhanced recommender services have the potential to support the retrieval process.

In a next step we plan to evaluate the proposed recommendation services [5] in a larger document assessment task where the services are utilized as query expansion mechanisms [8] and interactive services [15] in a realistic task scenario. Then we could measure indicators like actual task completion rates or goal satisfaction which need a quantitative basis.

However, a lot of research effort needs to be done to make more progress in coupling bibliometric-enhanced recommendation services with IR. The major challenge that we see here is to consider also the dynamic mechanisms which form the structures and activities in question and their relationships to dynamic features in scholarly information retrieval.

## 7. ACKNOWLEDGMENTS

My thanks go to all researchers in the study and my co-investigator Philipp Schaer for maintaining the online assessment tool. The work presented here was funded by DFG, grant no. INST 658/6-1 and grant no. SU 647/5-2.